\documentclass[conference]{IEEEtran}
\IEEEoverridecommandlockouts
\usepackage{amsmath,amssymb,amsfonts}
\usepackage{algorithmic}
\usepackage{graphicx}
\usepackage{textcomp}
\usepackage{xcolor}
\usepackage{comment}
\usepackage{cleveref}

\usepackage[style=ieee,backend=biber, url=false, doi=false, isbn=false]{biblatex}
\crefname{figure}{Fig.}{Figs.}
\Crefname{figure}{Fig.}{Figs.}
\begin{document}

\title{
Deliverable Frequency Regulation Reserve for VPPs via Transmission-Distribution Co-Simulation
}

\author{
    \IEEEauthorblockN{Ahmad Ali, Zhuorong Wang, 
    Youssef Abdelmoneim, and Hantao Cui}
    \IEEEauthorblockA{
        Department of Electrical and Computer Engineering \\
        North Carolina State University \\
        Raleigh, NC, USA \\
        \{aali27, zwang257, yaabdelm, hcui9\}@ncsu.edu
    }
}
\maketitle
\begin{abstract}
Aggregations of distributed energy resources can provide frequency regulation services.
However, device-level power headroom may not accurately represent the power deliverable at the transmission--distribution interface.
This paper proposes a bidirectional, storage-aware power-sharing framework for a virtual power plant (VPP).
Delivery factors obtained from the perturbations used to construct the voltage-sensitivity matrix account for feeder losses and voltage-dependent loads, while battery energy limits and state-of-charge weights are included in the allocation. 
The framework is evaluated using the IEEE 8500-node feeder and IEEE 14-bus system.
The proposed method tracks the feeder-head request within 2\% compared with 23-26\% over-delivery under the injection-headroom framework. 
It also sustains the requested power response 22\% longer before reaching the storage energy limit.
Moreover, communicating the updated VPP reserve to the transmission system operator reduces the post-transient integrated area control error by 80\% relative to fixed reserve registration.
\end{abstract}


\section{Introduction}
As the penetration of distributed energy resources (DERs) in the power grid grows, these devices are expected to provide ancillary services to the grid.
In such a setting, multiple small-scale DERs are managed via a central entity: a virtual power plant (VPP), which acts as an interface between the DERs and the transmission network \cite{pudjianto2007virtual}.
For the secondary frequency regulation (SFR) services, the transmission system operator (TSO) sends an automatic generation control (AGC) request to the VPP and observes the response at the interface with the distribution system operator (DSO).
The allocation of this request among individual DERs is managed internally by the VPP.
Consequently, the reserve registered with the TSO should represent the power that can be delivered at the interface.

The deliverable reserve of a VPP depends on resource availability, local network constraints, and network losses.
Resource availability is determined by the operating limits of each DER unit, such as the power rating, state of charge (SOC), and available headroom.
Local voltage limits further constrain the available headroom.
Moreover, the network losses change with the DER power injection and depend on the DER location, regulation direction, and feeder operating point.
Consequently, changing the power output of individual DERs does not necessarily lead to an equal change in power at the feeder head, and the VPP can overestimate or underestimate the actual reserve available to the TSO.

\subsection{Related Work}
Methods based on device-level headroom allocate regulation signals according to DER availability, operational constraints, and economic objectives \cite{chen2024real}.
Network-constrained approaches use power-flow models to enforce network constraints through online optimization \cite{8013070} or to find network-feasible injection limits \cite{9580648}.
However, headroom in \cite{chen2024real} and \cite{9580648} is expressed in terms of DER injections, which may differ from the power delivered at the feeder head.
The method in \cite{8013070} tracks the feeder-head power, but it requires online optimization.
Our previous work \cite{10556099} proposed a power-sharing method based on the voltage sensitivity matrix, which computes the network-constrained DER headroom and allocates an up-regulation request without online optimization.
Similarly, the registered reserve was based on device-level injections and did not account for the power delivered at the feeder head.
It also considered identical resources and a static operating point.
This work extends \cite{10556099} in three ways:
\begin{enumerate}
    \item \textbf{Interface-deliverable headroom:} Location-dependent delivery factors account for feeder losses and voltage-dependent loads between each DER and the transmission--distribution interface.
    
    \item \textbf{Bidirectional voltage-feasible reserve:} Upward and downward headroom account for resource limits, while a feasibility factor regulates the dispatch to satisfy the linearized voltage constraints.
    
    \item \textbf{Storage-aware allocation and evaluation:} SOC-dependent weights allocate the AGC request among energy-limited resources, and sequential transmission--distribution simulations evaluate the effects of inaccurate reserve registration on frequency and area control error.
\end{enumerate}

\section{Methodology}
\subsection{Voltage Sensitivities and Delivery Factors}
Let $\mathbf P_0$ and $\mathbf V_0$ be the DER active power injections and nodal voltage magnitudes at the current operating point, respectively.
For a perturbation $\Delta$ applied to the $i^{th}$ DER, one power flow solve yields the $i^{th}$ column of the VSM, $M_{ni}$ as:
\begin{align}
    M_{ni}=\frac{V_n(\mathbf P_0+\Delta\mathbf e_i)-V_n(\mathbf P_0)}
{\Delta}
\end{align}
where $\mathbf e_i$ is the unit vector associated with the $i^{th}$ DER.
The corresponding feeder-head delivery factor $s_i$ is computed as:
\begin{align}
    s_i=
\frac{P_h(\mathbf P_0)-P_h(\mathbf P_0+\Delta\mathbf e_i)}
{\Delta}
\end{align}
where $P_h$ is the active power imported at the feeder head.
Both DER injection and feeder import are defined as positive.
Therefore, $s_i$ represents the reduction in feeder-head import per unit increase in the power injection of the $i^{th}$ DER.
When the increase in DER injection reduces feeder losses, the change observed at the feeder head can be greater than the injected power, resulting in $s_i>1$. 
Similarly, $s_i<1$ when DER power adjustment is offset by feeder losses or voltage-dependent load.

For a signed DER adjustment vector $\mathbf r$, the voltage variation and delivered response, $P_{\mathrm{del}}$, are approximated as:
\begin{align*}
    \Delta\mathbf V\approx\mathbf M\mathbf r, \quad P_{\mathrm{del}}\approx\mathbf s^{\top}\mathbf r
\end{align*}
where $\mathbf s=[s_1,\ldots,s_N]^{\top}$. 

\subsection{Bidirectional Deliverable Headroom}
The VSM is used to compute the voltage-constrained candidate headroom vectors for upward and downward regulation:
\begin{align}\label{eq:net}
    \widehat{\mathbf P}^{+} = \mathbf M^{\dagger}
(\mathbf V_{\max}-\mathbf V_0), \quad \widehat{\mathbf P}^{-}
= 
\mathbf M^{\dagger}
(\mathbf V_0-\mathbf V_{\min})
\end{align}
where $\mathbf M^{\dagger}$ is the Moore--Penrose pseudo-inverse of the VSM.
In both $\widehat{\mathbf P}^{+}$ and $\widehat{\mathbf P}^{-}$, the negative values are clipped to zero.

The physical headroom depends on the resource type and its operating point.
For a PV unit, the upward and downward headroom are $P_{\mathrm{av},i}-P_{0,i}$ and $P_{0,i}$, respectively, where $P_{\mathrm{av},i}$ is the available PV power.
Thus, a PV unit operating at maximum power provides only downward regulation through curtailment.
Flexible loads can provide both upward and downward regulation by decreasing or increasing their power consumption within their operating band.
For a battery with power rating $P_i^r$, energy capacity $E_i$, SOC $\sigma_i$, and one-way efficiency $\eta_i$, the physical headroom over regulation horizon $H$ is \cite{8922750}:
\begin{align*}
P_{\mathrm{phys},i}^{+} &= \min\left(P_i^r-P_{0,i},\frac{\eta_i E_i(\sigma_i-\sigma_{\min})}{H}\right),\\
P_{\mathrm{phys},i}^{-} &= \min\left(P_i^r+P_{0,i},\frac{E_i(\sigma_{\max}-\sigma_i)}{\eta_i H}\right)
\end{align*}
The effective headroom of each DER is then obtained by combining its physical and the network limits:
\begin{align}
    P_{\mathrm{eff},i}^{\pm} = \min\left( \widehat P_i^{\pm}, P_{\mathrm{phys},i}^{\pm} \right)
    \label{eq:eff_headroom}
\end{align}

However, \eqref{eq:net} is a least-squares approximation, which implies 
$\mathbf M\widehat{\mathbf P}^{+} \neq \mathbf V_{\max}-\mathbf V_0$. 
So the headroom obtained using \eqref{eq:eff_headroom} may lead to voltage-limit violation.
To ensure $\mathbf V_{\min} \leq \mathbf{V}_0 + \mathbf{M}(\pm \mathbf{P}_{\mathrm{eff}}^{\pm}) \leq \mathbf{V}_{\max}$, a voltage-feasibility scaling factor, $\lambda$ is introduced.
For the adjustment vector $\mathbf r$, $\lambda$ is defined as:

\begin{align}
    \lambda(\mathbf r)=
\min\left(
1,
\min_{n:(\mathbf M\mathbf r)_n>0}
\tfrac{V_{\max}-V_{0,n}}{(\mathbf M\mathbf r)_n},
\min_{n:(\mathbf M\mathbf r)_n<0}
\tfrac{V_{0,n}-V_{\min}}{-(\mathbf M\mathbf r)_n}
\right)
\label{eq:volt_feas_fac}
\end{align}
To ensure existing violations are not worsened, the factor is set to zero when a node that limits the adjustment is already outside its voltage limit.
The upward and downward reserves registered with the TSO are then computed as:
\begin{align}
    R^{\pm} = \lambda\left(\pm\mathbf P_{\mathrm{eff}}^{\pm}\right) \mathbf s^{\top}\mathbf P_{\mathrm{eff}}^{\pm}.
\end{align}

Therefore, \(R^{+}\) and \(R^{-}\) represent the voltage-feasible regulation power that can be delivered at the TSO–DSO interface.

\subsection{SOC-Weighted AGC Allocation}
For a signed AGC request $P_{\mathrm{agc}}$, the upward headroom is used when $P_{\mathrm{agc}}>0$, and the downward headroom is used when $P_{\mathrm{agc}}<0$.
The power adjustment assigned to the $i^{th}$ DER is computed based on its effective headroom and the participation factor as:
\begin{align}
    r_i=
\operatorname{sgn}(P_{\mathrm{agc}})
P_{\mathrm{eff},i}^{\pm}
\min(1,\alpha w_i^{\pm})
\end{align}
where $w_i^{\pm}$ is the SOC-based priority weight and $\alpha$ is the participation factor.
The participation factor is selected as:
\begin{align}
    \sum_{i=1}^{N}s_i|r_i|
=
\min\left(
|P_{\mathrm{agc}}|,
\sum_{i=1}^{N}s_iP_{\mathrm{eff},i}^{\pm}
\right)
\end{align}

Thus, the allocation is based on the power delivered at the feeder head rather than the sum of the DER power adjustments.
When no DER is saturated, $\alpha = |P_{\mathrm{agc}}|/(\sum_{i=1}^{N}s_iP_{\mathrm{eff},i}^{\pm} w_i^{\pm})$.
The DER $i$ reaches its maximum headroom when $\alpha = 1/w_i^{\pm}$.
When a DER saturates, its contribution becomes fixed at $P_{\mathrm{eff},i}^\pm$.
Therefore, $\alpha$ can be obtained directly from the breakpoints $1/w_i^{\pm}$ without solving an optimization problem.

The participation weights are set to one for PV and flexible-load resources.
For batteries, they are defined as
\[
w_i^{+}
=
\left[
\frac{\sigma_i-\sigma_{\min}}
{\sigma_{\max}-\sigma_{\min}}
\right]_{w_{\min}}^{1},
\!
w_i^{-}
=
\left[
\frac{\sigma_{\max}-\sigma_i}
{\sigma_{\max}-\sigma_{\min}}
\right]_{w_{\min}}^{1}
\]
where $[\cdot]_{w_{\min}}^{1}$ limits the weight to $[w_{\min},1]$. 
A battery with higher SOC receives a larger share of an upward request, whereas a battery with lower SOC receives a larger share of a downward request.
The allocated power is checked using the voltage-feasibility factor in \eqref{eq:volt_feas_fac}:
\[
\mathbf r^{*}=\lambda(\mathbf r)\mathbf r,
\qquad
P_{\mathrm{del}}=\mathbf s^{\top}\mathbf r^{*}.
\]
When the full request cannot be delivered, the difference \(P_{\mathrm{agc}}-P_{\mathrm{del}}\) is reported as a shortfall. 
After every AGC interval $\Delta t$ the SOC of each battery is updated.

\begin{figure*}[!t]
    \centering
    \includegraphics[width=0.9\linewidth]{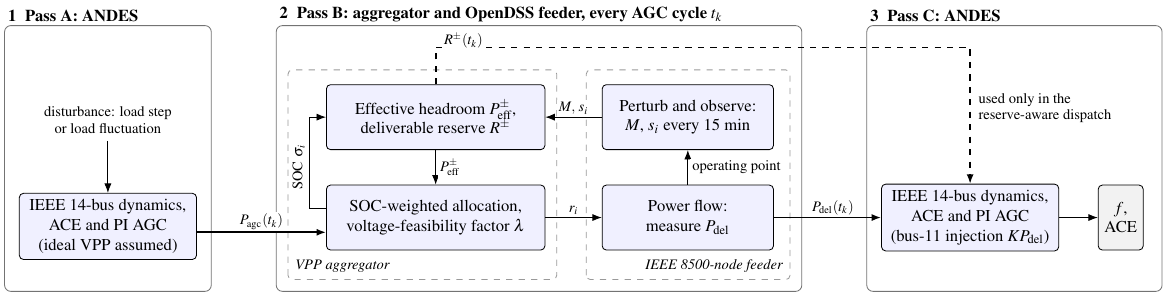}
    \caption{Sequential evaluation framework of transmission-distribution co-simulation.}
    \label{fig:main_idea}
\end{figure*}
\section{Sequential Transmission-Distribution Co-simulation}
The proposed scheme is evaluated using a three-pass transmission-distribution co-simulation framework shown in \Cref{fig:main_idea}.
In Pass A, the IEEE 14-bus system is simulated in ANDES \cite{9169830} with five synchronous generators, TGOV1 governors, and an area control error (ACE) calculator.
The total AGC command is obtained using a proportional-integral controller.
A fraction $\rho$ of the total AGC command is assigned to the VPP, while the remaining command is distributed equally among the conventional generators.
The VPP command is recorded every $\Delta t=4$~s by assuming that the VPP delivers the requested power.

In Pass B, the recorded command is applied to the IEEE 8500-node feeder in OpenDSS.
At each AGC cycle, the DER operating points are updated, the AGC request is allocated using the framework in Section II, and the resulting change in feeder-head power is recorded as \(P_{\mathrm{del}}\).
The voltage sensitivities and delivery factors are updated every 15~min.

In Pass C, the transmission simulation is repeated using the delivered power obtained from Pass B to evaluate the resulting frequency and ACE.

The sequential procedure approximates the closed-loop transmission–distribution interaction.
If the VPP delivers a fraction $\epsilon$ more than the requested power, the steady-state power balance $(1-\rho)u+(1+\epsilon)\rho u=\Delta P$ gives $u=\Delta P/(1+\rho\epsilon)$, where $\Delta P$ is the disturbance.
Thus, the difference between the recorded and closed-loop VPP requests is approximately $\rho\epsilon$.
This difference is below 1\% for the proposed method and the linear programming (LP)-based benchmark, and approximately 7\% for the injected-headroom method used in \cite{10556099}.

\section{Case Studies and Results}
\subsection{Simulation Setup}
The VPP consists of 200 DERs randomly placed at load buses.
The DER units consist of 100 PV units rated at 3 kW, 60 batteries rated at 5~kW/3~kWh, and 40 flexible loads rated at 4~kW with a $\pm2$~kW flexibility band.
The PV units initially operate at their maximum available power.
The battery SOC is uniformly distributed from 0.3--0.9, with allowable SOC range of $[0.1,0.9]$.
The battery one-way efficiency is $\eta=0.95$ and the minimum weight is $w_{\min}=0.1$.
The regulation horizon is $H=15$ min.
The voltage limits are $[0.917,1.058]$~pu. 
During the 30-minute simulation, the feeder load increases linearly from 0.95 to 1.05 of its nominal value, and a cloud transient reduces the PV irradiance between 10 and 14~min.

The IEEE 14-bus system supplies a total load of 224 MW.
The AGC gains are $K_P=K_I=0.1$, the frequency bias is $-16.3$ MW/0.1 Hz, and the VPP participation factor is $\rho=0.3$.
The AGC interval is 4~s, and $K=12$ identical feeders are represented behind bus 11. 

Two cases are studied.
In the first case, a 10~MW load increase is applied at bus 3, resulting in an upward regulation.
In the second case, an autoregressive load variation with standard deviation of 1.3\% produces a bidirectional AGC signal.
The proposed allocation method is compared against the injected-headroom method, and the LP benchmark described in Section IV-B.

\subsection{LP Benchmark}
An LP-based optimal allocation rule is used as a benchmark.
The formulation is based on \cite{10556099} but is extended to include the delivery factors derived in Section II-A.
Let $\zeta=\mathrm{sgn}(P_{\mathrm{agc}})$ and let $d$ denote the portion of the AGC request that cannot be delivered.
At each AGC cycle, the LP solves
\begin{equation*}
\begin{aligned}
\min_{\mathbf r,d}\quad
& d+\gamma\zeta\mathbf{1}^{\top}\mathbf M\mathbf r \\
\mathrm{s.t.}\quad
& \zeta\sum_{i=1}^{N}s_i r_i+d
=|P_{\mathrm{agc}}|,\\
& \min(\mathbf V_{\min},\mathbf V_0)\leq\mathbf V_0+\mathbf M\mathbf r
\leq \max(\mathbf V_{\max},\mathbf V_0),\\
& 0\leq\zeta r_i\leq P_{\mathrm{eff},i}^{\pm},
\quad i=1,\ldots,N,\\
& d\geq0 .
\end{aligned}
\end{equation*}

The voltage constraint ensures that nodes initially within the prescribed limits remain within them, while existing violations are not worsened.
The constraint on $r_i$ prevents individual DERs from operating opposite to the regulation direction.
The second term in the objective minimizes the aggregate voltage change, and the coefficient $\gamma$ ensures that minimizing the power shortfall is the main objective.
The resulting problem is solved using HiGHS.

\subsection{Time-Varying Deliverable Reserve}
\begin{figure}
    \centering
    \includegraphics[width=0.85\linewidth]{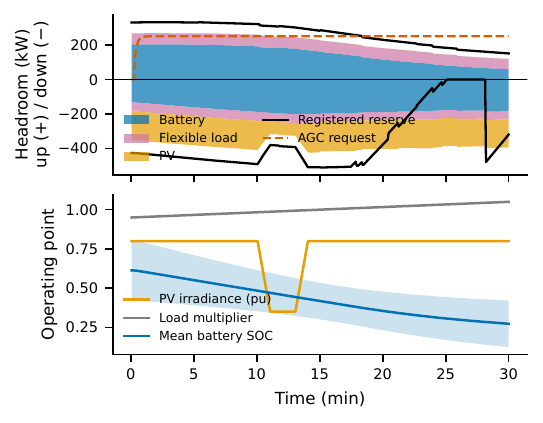}
    \caption{Effective headroom by resource in both
directions, registered reserve, AGC request (top); load multiplier, irradiance,
and battery SOC (bottom).}
    \label{fig:der_reserve}
\end{figure}
\Cref{fig:der_reserve} shows the effective headroom of the DER units during the 30-minute load-step scenario in both directions, and the AGC command.
Initially, the batteries and flexible loads provide 204 kW and 64 kW of up-regulation headroom, respectively.
The PV units are operating at their maximum power points and hence do not contribute in this direction.
After accounting for the delivery factors, the registered up-regulation reserve is 331~kW, which exceeds the requested 251~kW.
The registered reserve decreases to 286 kW at $t=15$ min and 152 kW at $t=30$ min as the batteries discharge.

The initial down-regulation reserve is 427 kW. It reduces by approximately 150 kW
during the cloud transient between 10 and 14 min.
The down-regulation reserve collapses to zero between 25 and 28 min, when the load ramp brings up to 18 nodes of the base case below the 0.917 pu limit.
Any further reduction of injection would worsen the violation.
A subsequent regulator tap change restores the available voltage margin and the down-regulation reserve.
These results show that the deliverable reserve varies with the battery energy state, PV availability, and feeder operating conditions. 

\subsection{Response Ratio at the Substation}
\begin{figure}
    \centering
    \includegraphics[width=0.85\linewidth]{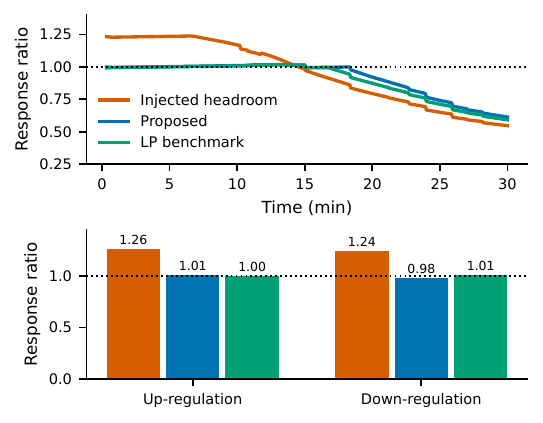}
    \caption{Response ratio at the feeder head for the three allocation methods:
per AGC cycle in the load-step scenario (top) and over the up- and down-regulation in the load-fluctuation scenario (bottom).}
    \label{fig:der_rr}
\end{figure}
\Cref{fig:der_rr} compares the response ratio of the three allocation schemes.
The response ratio is defined as the power delivered at the feeder head divided by the requested power.
It is evaluated at each AGC cycle for the load-step scenario and over the total regulation energy in each direction for the bidirectional scenario.

In the load-step scenario, the injected-headroom method initially delivers 23--24\% more power than requested.
This additional battery usage causes the injected-headroom method to become energy-limited at $t=15.1$ min.
In comparison, the proposed method maintains the response ratio within 2\% of unity until it becomes energy-limited at $t=18.4$ min,
while the LP method becomes energy-limited at $t=17.2$ min.
The LP does not consider the SOC in its objective, and hence it depletes the batteries earlier.

For the bidirectional case, the injected-headroom method produces response ratios of $1.26$ and $1.24$ for up- and down-regulation, respectively.
It also causes voltage violation in 16 of the 450 AGC cycles.
In contrast, the proposed method produces response ratios of 1.01 and 0.98, while the LP produces ratios of 1.00 and 1.01.
The down-regulation deficit of the proposed method occurs during the three-minute interval during which the registered down-regulation reserve is zero.
Excluding this interval, its down-regulation response ratio is 1.01.
The LP continues to deliver down-regulation during this interval because it selects individual DERs that do not reduce the voltage at the limiting nodes, whereas the proposed method applies a common feasibility factor to all units.

The LP benchmark does not cause any voltage violation, whereas the proposed method produces a small violation in three consecutive cycles at one node.
The voltage is \(0.4\text{--}2.7\times10^{-5}\) pu below the lower limit. 
These violations were due to stale voltage sensitivities following regulator tap changes.
This can be avoided by updating the sensitivities after regulator operations or by including a small voltage margin in \eqref{eq:volt_feas_fac}.

To evaluate the role of the feasibility factor $\lambda$, an additional test is conducted with static AGC requests of 300--500~kW on 200 identical DERs, each operating at 1 kW with 3 kW of available headroom.
The injected-headroom method causes 9 voltage violations at a request of 300 kW and 19 violations at 400 kW.
The proposed method applies feasibility factors of 0.83 and 0.67 at requests of 400 and 500 kW, respectively, and delivers 331 kW without violating any voltage limits.
Since the LP method selects the DERs individually, it delivers 397 kW and 458 kW for the same requests.

\subsection{Frequency and ACE Response}
\begin{figure}
    \centering
    \includegraphics[width=0.85\linewidth]{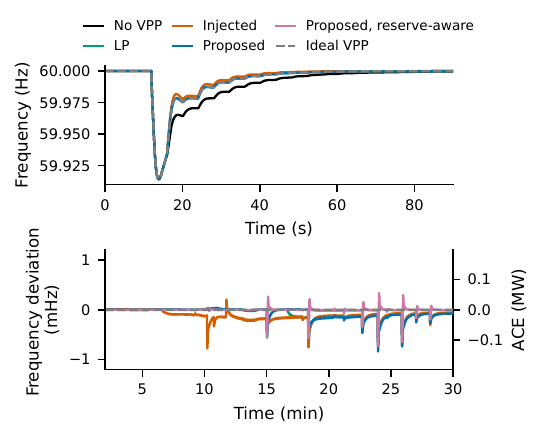}
    \caption{System frequency in the step scenario: the disturbance transient (top)
and the deviation after the transient, equivalent ACE is on the right axis (bottom).}
    \label{fig:freq_ace}
\end{figure}

\Cref{fig:freq_ace} shows the system frequency for the load-step scenario.
Without VPP participation, the synchronous generators supply the entire 10~MW response, the ACE settles in 37~s, and the integrated absolute ACE is 155 MW~s.
With an ideal VPP response, the settling time decreases to 
29~s and the integrated ACE decreases to 108 MW~s.
Since the requested VPP power is within the available reserve, the proposed method initially reproduces the ideal response.
After approximately 18 min, the battery energy limit is reached and the VPP response decreases from 3.0~MW to 1.9~MW.
When a fixed reserve is registered to the TSO, the governors continue to receive only 70\% of the AGC command and the remaining power imbalance appears in the ACE. 
The governor response ultimately increases to counter the power imbalance.

The proposed, injected-headroom, and LP methods produce integrated ACE values of 128 MW~s, 131 MW~s, and 129 MW~s, respectively.
Since the injected-headroom method over-delivers the requested power, it shortens the frequency recovery by approximately one second.
However, this method causes the batteries to drain approximately three minutes earlier than the proposed method.
Its post-transient integrated ACE is therefore 33 MW\,s, compared with 20 MW\,s for the proposed method. 

A reserve-aware dispatch is also evaluated where the TSO receives the updated deliverable reserve from the VPP at every AGC cycle. 
This reduces the post-transient integrated ACE from 20 MW~s to 4 MW~s.
In the load-variation case, VPP participation reduces the RMS ACE from 1.33 MW to 1.20 MW.
The RMS ACE obtained using the proposed method and the LP remains within 0.3\% of the ideal VPP result.

\subsection{Computation Time}
Table~\ref{tab:time} presents the computation time as the number of DERs increases from 100 to 600.
The computation of VSM, delivery factors, and effective headroom are common to both methods.
The headroom computation requires less than 2 ms per cycle, while the update of VSM and delivery factors requires 1.4 s for 200 units and 5.0 s for 600 units once every 15 min.
The allocation time of the proposed method remains below 1 ms, whereas the LP requires approximately 1~s for 600 units. 
Since no voltage limit is active, both methods deliver the same power in these cases.
\begin{table}[!t]
    \centering
    \caption{Allocation time per AGC cycle}
    \label{tab:time}
    \begin{tabular}{ccc}
    \hline
    Number of DERs & Proposed Method (ms) & LP Benchmark (ms)\\
    \hline
    100 & 0.14 & 9\\
    200 & 0.16 & 35\\
    400 & 0.27 & 356\\
    600 & 0.40 & 1009\\
    \hline
    \end{tabular}
\end{table}

\section{Conclusion}
A bidirectional power-sharing framework for quantifying and allocating the regulation reserve of a VPP is proposed.
The framework uses delivery factors to compute the DER headroom as the power available at the TSO--DSO interface.
The main conclusions are:
\begin{enumerate}
    \item VPP headroom computation based on device-level availability can over- or underestimate the power delivered at the feeder head.
    In the studied case, such a rule over-delivers the requested power by 23--26\%, leading to early resource depletion and a larger post-transient ACE.
    \item The injected-headroom method causes up to 19 voltage violations in the stressed case, whereas the proposed feasibility factor prevents these violations under the linear voltage model.
    \item Compared with the LP optimum, the proposed method achieves similar tracking when no voltage limit is active.
    Under stressed conditions, it delivers 72--100\% of the LP optimum at less than 2\% of the computational cost.
    \item Communicating the updated deliverable reserve to the TSO reduces the post-transient integrated absolute ACE from 20 MW~s to 4 MW~s.
\end{enumerate}

\section*{Acknowledgment}
Claude (Anthropic) was used for language editing and manuscript polishing.

\printbibliography

@INPROCEEDINGS{10556099,
  author={Ali, Ahmad and Cui, Hantao and Wang, Wenbo and Fang, Xin},
  booktitle={2024 IEEE/PES Transmission and Distribution Conference and Exposition (T\&D)}, 
  title={Power Sharing-Based Framework for Allocating Automatic Generation Control in Distributed Energy Resources}, 
  year={2024},
  volume={},
  number={},
  pages={1-5},
  doi={10.1109/TD47997.2024.10556099}}

@ARTICLE{9169830,
  author={Cui, Hantao and Li, Fangxing and Tomsovic, Kevin},
  journal={IEEE Transactions on Power Systems}, 
  title={Hybrid Symbolic-Numeric Framework for Power System Modeling and Analysis}, 
  year={2021},
  volume={36},
  number={2},
  pages={1373-1384},
  doi={10.1109/TPWRS.2020.3017019}}

@article{chen2024real,
  title={Real-time operation strategy of virtual power plants with optimal power disaggregation among heterogeneous resources},
  author={Chen, Qixin and Lyu, Ruike and Guo, Hongye and Su, Xiangbo},
  journal={Applied Energy},
  volume={361},
  pages={122876},
  year={2024},
  publisher={Elsevier}
}

@article{pudjianto2007virtual,
  title={Virtual power plant and system integration of distributed energy resources},
  author={Pudjianto, Danny and Ramsay, Charlotte and Strbac, Goran},
  journal={IET Renewable Power Generation},
  volume={1},
  number={1},
  pages={10--16},
  year={2007},
  publisher={IET}
}

@ARTICLE{8013070,
  author={Dall’Anese, Emiliano and Guggilam, Swaroop S. and Simonetto, Andrea and Chen, Yu Christine and Dhople, Sairaj V.},
  journal={IEEE Transactions on Power Systems}, 
  title={Optimal Regulation of Virtual Power Plants}, 
  year={2018},
  volume={33},
  number={2},
  pages={1868-1881},
  doi={10.1109/TPWRS.2017.2741920}}

@ARTICLE{9580648,
  author={Nazir, Nawaf and Almassalkhi, Mads},
  journal={IEEE Transactions on Power Systems}, 
  title={Grid-Aware Aggregation and Realtime Disaggregation of Distributed Energy Resources in Radial Networks}, 
  year={2022},
  volume={37},
  number={3},
  pages={1706-1717},
  doi={10.1109/TPWRS.2021.3121215}}

@ARTICLE{8922750,
  author={Rosewater, David M. and Copp, David A. and Nguyen, Tu A. and Byrne, Raymond H. and Santoso, Surya},
  journal={IEEE Access}, 
  title={Battery Energy Storage Models for Optimal Control}, 
  year={2019},
  volume={7},
  number={},
  pages={178357-178391},
  doi={10.1109/ACCESS.2019.2957698}}
\end{document}